\documentclass[acmsmall, authorversion, nonacm=true]{acmart}

\acmPrice{}
\acmYear{}
\acmDOI{}

\acmConference[ProLaLa 2023]{Workshop on Programming Languages and the Law 2023}{15 Jan 2023}{Boston, Mass., USA}
  
\usepackage{amsmath,amsfonts}
\usepackage{graphicx}
\usepackage{textcomp}
\usepackage{comment}
\usepackage{multirow}
\usepackage{caption}
\usepackage{subcaption}

\usepackage{makecell}
\usepackage{tabularx}
\usepackage{eurosym}
\usepackage{wrapfig}
\usepackage{csquotes}

\usepackage{hyperref}
\usepackage{url}

\newcommand{\owner}{\textit{owner}}

\newcommand{\damage}{\textit{Compensation}}

\newcommand{\SPA}{\textit{SPA}}

\let\ACMmaketitle=\maketitle
\renewcommand{\maketitle}{\begingroup\let\footnote=\thanks \ACMmaketitle\endgroup}

\begin{document}

\title[ContractCheck]{Formal Modeling and Analysis of Legal Contracts using \textit{ContractCheck}\\
-- Extended Abstract --}

\titlenote{Accepted for presentation at: Workshop on Programming Languages and the Law 2023, ProLaLa23, 15 Jan 2023, Boston, Mass., USA.}


\author{Alan Khoja}
\email{Alan.Khoja@uni.kn}
\orcid{}
\affiliation{%
  \institution{University of Konstanz, Dept.\ of Law}
  \country{Germany}
}

\author{Martin K\"olbl}
\email{Martin.Koelbl@uni.kn}
\authornote{The majority of contributions of this author were 
made while affiliated with the Unviersity of Konstanz, Germany.}
\orcid{}
\affiliation{%
  \institution{CertiK}
  \country{USA}
}

\author{Stefan Leue}
\email{Stefan.Leue@uni.kn}
\authornote{Presenting author.}
\orcid{0000-0002-4259-624X}
\affiliation{%
  \institution{University of Konstanz, Dept.\ of Computer Science}
  \country{Germany}
}

\author{R\"udiger Wilhelmi}
\email{Ruediger.Wilhelmi@uni.kn}
\orcid{}
\affiliation{%
  \institution{University of Konstanz, Dept.\ of Law}
  \country{Germany}
}


\keywords{Legal contracts, logic and law, consistency, SMT, ContractCheck}

\begin{abstract}
We describe a method and tool called \textit{ContractCheck}~\cite{DBLP:conf/spin/KhojaKLW22} that allows
for the consistency analysis of legal contracts, in particular Sales 
Purchase Agreements (SPAs). The analysis relies on an encoding of the premises 
for the execution of the 
clauses of an SPA as well as the proposed consistency constraints 
using decidable fragments of first-order logic. Textual SPAs are first 
encoded in a structured natural language format, called blocks. 
\textit{ContractCheck} interprets these blocks and constraints and
translates them in first-oder logic assertions. It then invokes a
Satisfiability Modulo Theories (SMT) solver in order to establish 
the executability of a considered contract by either providing a satisfying model, 
or by providing evidence of contradictory 
clauses that impede the execution of the contract. We illustrate 
the application of \textit{ContractCheck} and conclude by proposing
directions for future research.
\end{abstract}

\maketitle

\section{Introduction}

Legal contracts are an essential basis for the conduct of business.
The particular type of legal contract that we shall focus on is 
a share purchase agreement (SPA) which regulates the sale of a company.
Contracts, and especially SPAs, are often very long and complex. 
In addition, a large number of persons are often involved in the drafting.
Moreover, the negotiations and the drafting may take long 
and comprise a large number of amendments of the draft.
Length, complexity and frequent changes make a contract prone
to errors and inconsistencies, such as references being wrong,
missing essentials or unfulfillable claims. In this paper we shall
focus on semantic errors in SPAs, in particular unfullfillable claims.

Building on our experience in modeling and analysis for complex software systems engineering, we are 
presenting a method to formalize SPAs using decidable fragments of 
first order logic, as well as a tool called \textit{ContractCheck}~\cite{DBLP:conf/spin/KhojaKLW22},
which automatically checks consistency conditions on the formalization
using Satisfiability Modulo Theory (SMT) solving~\cite{barrett2018satisfiability}.
We choose this approach since full automation of the analysis promises 
easy acceptance of the approach in the tool infrastructure of the legal domain, in particular compared
to interactive theorem proving.

{\centering
\begin{figure}
\begin{subfigure}{.5\textwidth}
  \centering
  \includegraphics[width=.6\linewidth]{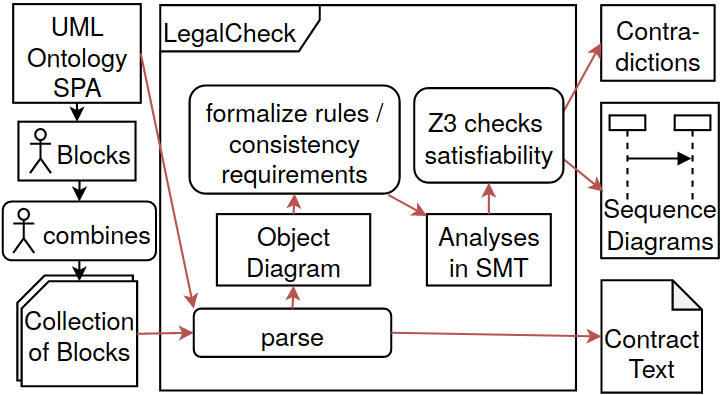}
  \caption{The \textit{ContractCheck} Approach~\cite{DBLP:conf/spin/KhojaKLW22}.}
  \label{fig:ap_stm}
\end{subfigure}%
\begin{subfigure}{.5\textwidth}
  \centering
  \includegraphics[width=.8\linewidth]{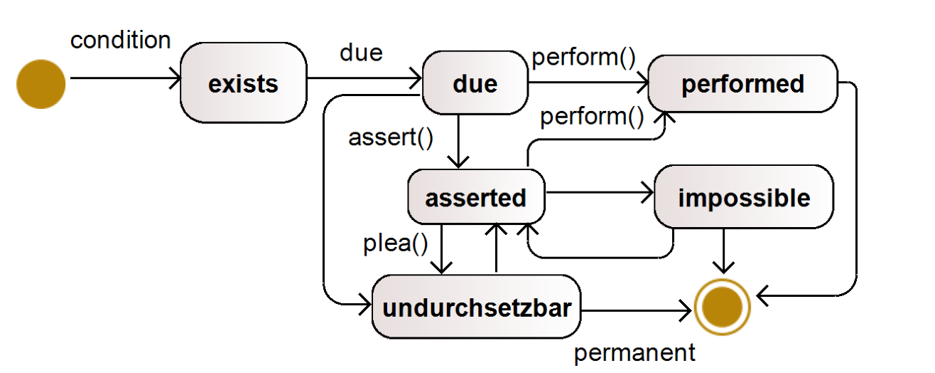}
  \caption{State machine model of the object claim~\cite{DBLP:conf/spin/KhojaKLW22}.}
  \label{fig:claim}
\end{subfigure}%
\end{figure}
}

Figure~\ref{fig:ap_stm} presents the the main workflow structure of 
\textit{ContractCheck}~\cite{DBLP:conf/spin/KhojaKLW22}. In order to identify the concepts that the 
analysis needs to reason about, we first elicited an ontology in that work.
The user of \textit{ContractCheck} will 
map the clauses from the textual form of the SPA to a combination of pre-defined structured 
language blocks. This could, for instance, happen inside a contract generation 
tool. \textit{ContractCheck} parses this collection, translates it into 
an UML Object Diagram used as an internal representation of the considered
contract instance, and translates this internal representation 
into a set of logical assertions in decidable fragments of quantifier-free first order 
logic (linear integer arithmetic, uninterpreted functions, equality).
This set of assertions encodes the different clauses in the contract,
in particular the logical premises for their performance,
as well as the applicable consistency requirements. It is then
submitted to an SMT solver, in our case Z3~\cite{deMBjo08}, which checks the constraints 
for satisfiability and either returns a satisfying model or an unsatisfiability 
core. The results of the SMT solving are depicted and presented to the user.

In this paper, we restrict ourselves to considering SPAs, since  
they are characterized by little reliance on implicit legal facts implied by 
legal dogmatics, which are typically waived in SPAs. 
While we develop \textit{ContractCheck} using an SPA under German law, we contend
that this does not preclude its adaptation to contracts with a different subject 
matter or under other jurisdictions.

\paragraph{Related Work.}
Historically, the logical modeling of law was proposed by G.W.\ Leibnitz as early as in the 
17th centrury~\cite{Arm.2014}. More recently, deontic and defeasible logics have been
developed with an expressiveness designed to represent legal
facts~\cite{CasMai07,PriSch12,BalBroBru09}. Domain-specific modeling languages
for contracts have been defined~\cite{PriSch12,GarMerSch10,CamSch17} and the analysis 
of these languages using model checking has been
proposed~\cite{PacPriSch07,GarMerSch10,GorMerSch11,CamSch17}.
Closest to our approach is the formal contract specification language Symboleo~\cite{9218159}. However, it focuses
on providing executable contract  specifications, not on formal analysis.
The objective of our work is to analyze both the dynamic execution of 
contracts as well as to reason about problem-domains, for instance using integers
representing purchase prices, restitutions and pricing. Also, since we are not
interested in analyzing infinite executions, we see no benefit in using Linear
Time Temporal Logic. 
Due to its fully automatic nature, 
the SMT-solving framework that we use in our approach
appears to be a very suitable environment for the implementation of our analyses.

\section{Modeling}

The quintessential concept of a contract is that of mutual claims, i.e., the right
of a legal subject to demand an act or omission from another legal subject.
In relation to a particular claim in an SPA, the debtor, who owes its performance,
is either the purchaser or the seller. The respective other party is the creditor.
Claims can become due, and can be asserted and performed. Claims have a state, and 
a (potentially incomplete) state machine for the object claim, a depiction of which can be found in 
Figure~\ref{fig:claim}. State changes in claims contribute to state changes of
a contract in execution. They can, hence, be considered the "computation steps" 
during the execution of a contract, to draw an analogy with computing.

\paragraph{Bakery SPA Example.}
As a running example, we use an SPA for the fictitious sale of
a bakery from the seller \texttt{Eva} to the purchaser \texttt{Chris} to illustrate
the modeling and analysis in \textit{ContractCheck}.
In the bakery SPA, Chris agrees to pay the purchase price of \euro $40,000$  (\texttt{PayClaim}), 
and Eva agrees to give Chris ownership of the bakery \texttt{Shares} \texttt{Bakery} (\texttt{TransferClaim}).
Both claims are due at the agreed closing $28$ days after signing.
In case the pretzel bakery is not transferred or the purchase price 
is not paid at closing, then the other contracting party may 
withdraw from the purchase contract.
The contract comes into force on the day of the signature of the
contracting parties (Signing) and usually ends ("expires")
when the purchase object is handed over (Closing).
In addition to the \textit{primary claims}, 
Eva guarantees in a \texttt{WarrantyClaim} called \texttt{PretzelWarranty}
that the bakery can bake $10,000$ pretzels a day.
If the bakery in the example cannot meet this promised performance, 
then the warranty is breached and Chris has to assert this breach 
within the \texttt{DueDate} of $14$ days of closing.
In case of an assertion, Eva has to make good within $28$ days
because of the \texttt{PerformanceClaim} \texttt{Claim1}, 
otherwise she has to pay within $14$ days a compensation of \euro $1,000$
per $100$ of pretzels that cannot be baked, as specified in the the \texttt{CompensationClaim}
\texttt{Claim2}. Notice that 
\texttt{Claim2} has a minimal compensation of \euro $1000$.
Any claim under the warranty has a \texttt{Limitation} of $42$ days of closing. 
We currently use an auxiliary construction that environment facts are 
encoded as clauses in the contract. In particular, we assume that a bank named \texttt{Bank} 
has ownership  by way of security of the shares of the bakery.

The different clauses from the contract are handed over to \textit{ContractCheck} using
parameterized structured English text blocks.
\textit{ContractCheck} parses these blocks, and translates them to the relevant logical encoding.
We now present a few examples of this encoding. A more detailed and principled discussion
of the encoding as well as more encoding examples can be found in~\cite{DBLP:conf/spin/KhojaKLW22}.

\paragraph{Ownership.}
SPAs regulate the transfer of property. This requires that the debtor of a 
transfer claim is actually owner of the object to be transferred.
We define a relation $\mathcal{PR}$ and add to it a tuple $(p,o)$ for every
instance of \texttt{PropertyRight}.
We formalize the property rights by an uninterpreted function 
$\owner(\textit{Object}):\textit{Person}$.
$\phi_{\owner}$ represents the property relations stated in the SPA:
$\phi_{\owner} = \bigwedge_{(p,o)\in \mathcal{PR}} \owner(o)=p.$
For the Pretzel Bakery SPA, for instance, it holds that  
$\phi_{\owner}^B\equiv\owner(\textit{Bakery})= \textit{Bank}.$

\paragraph{Claims.}
We use integer arithmetic and uninterpreted functions to express
claims that entail transfer of property or financial resources.
For a claim $C$, the integer $d_C = -1$ indicates a state of the
contract, in which the claim has not (yet) been performed. A 
value $d_C = n$ for $n$ a positive integer asserts that the claim
is performed on day $n$ after signing. 

The transfer claim that Eva needs to hand over the 
shares for the Bakery to Chris on closing day, which is day
28 after signing, is formalized by 
$\phi_{\textit{TransferClaim}} \equiv \; (-1= d_u)\vee
((28\leq d_u)\Rightarrow
(\owner(\textit{Bakery})=\textit{Eva})).$
Notice that $\owner(\textit{Bakery})=\textit{Eva}$ expresses the logical 
premise for the performance of \textit{TransferClaim}.
The fact that Chris is obliged to pay the PurchasePrice 
is expressed by the PayClaim:
$\phi_{\textit{PayClaim}} \equiv
(-1= d_z)\vee((28\leq d_z)\Rightarrow
\owner(\textit{PurchasePrice})=\textit{Chris})).$
The transfer is performed on a day $d_z$ and is due
on day $28$. Restitution claims constraining the breach of any
of these two claims are also defined~\cite{DBLP:conf/spin/KhojaKLW22}.

\paragraph{Warranties.}
The modeling framework also allows for the handling of warranties when claims are breached.
If the \textit{warranty claim} \textit{PretzelWarranty} is breached,
then Chris notifies this breach on a day $d_g\geq 0$.
$d_g=-1$ means that there is no indication of a non-performance and 
therefore the warranty condition $\textit{Pretzels}=10,000$ is met.
The formalization for the \textit{PretzelWarranty} is 
$\phi_{\textit{PretzelWarranty}}\equiv\;
(d_g=-1\Rightarrow Pretzels=10000)\vee
(28\leq d_g\leq 28+14).$
In the bakery SPA, the warranty has
the consequence \textit{Claim1}, so that the seller can subsequently perform the pretzel guarantee on a date $d_n$: 
$\phi_{\textit{Claim1}}\equiv
(d_n=-1) \vee (d_g< d_n\leq d_g+28 \Rightarrow
Pretzels=10000).\label{eq:folge1}$
On the other hand, it may be more advantageous for the debtor
to pay a compensation $l_s$ on a date $d_s$ ("efficient breach").
The value of $l_s$ is constrained by the formula $\phi_\damage^s$~\cite{DBLP:conf/spin/KhojaKLW22}.
If no compensation occurs ($d_s=-1$), then $l_s$ is $0$.
The formalization of the compensation claim \textit{Claim2} is
$\phi_{\textit{Claim2}}\equiv \; \phi_\damage^s \wedge
((d_s=-1 \Rightarrow l_s= 0) \;\vee 
(d_g< d_s\leq d_g+ 28+14)).$

\paragraph{Contract Execution.}
%
The overall encoding of the claims in the bakery SPA is 
given by the transfer, pay and warranty claims, where each is disjoined with
the the respective secondary claim: 
$\phi_\SPA^B\equiv \phi_{\owner}^B\wedge (\phi_{\textit{TransferClaim}}\vee
\phi_{\textit{Res.Purchaser}})\wedge(\phi_{\textit{PayClaim}}\;\vee
\phi_{\textit{Res.Seller}})\wedge
(\phi_{\textit{PretzelWarranty}}\vee \phi_{\textit{Claim1}}\vee \phi_{\textit{Claim2}}).$
While the contract parties are free to choose whether to perform or 
breach a claim, it may be desirable to determine whether 
an execution of the contract, in which primary claims are performed, is possible at all.
In order to analyze this condition we introduce soft constraints representing secondary claims that the 
SMT solver strives to satisfy a minimal subset from in a satisfying model:
%
$\phi_{\textit{soft}}^B\equiv \;
d_u\geq 0\wedge d_z\geq 0\wedge d_{\textit{Res.Purchaser}}=-1
\; \wedge 
d_{\textit{Res.Seller}}=-1\wedge d_g=-1 \;\wedge\; d_n=-1$
%
The bakery SPA is formalized by the constraint 
$\phi_\SPA^B \wedge \phi_{\textit{soft}}^B$. Each satisfying model
of this constraint represents a possible execution of the bakery SPA.

\section{Analysis}

We define two types of analysis. Analysis I determines the executability
of every claim. Let $d_c$ express the constraints on primary claim $c$, which has to 
perform on date $d_c$, and let $\phi_{\owner}$ express the property relation for
this claim, then the satisfiability of 
$\Phi_c\equiv\phi_{\owner}\wedge \phi_c\wedge d_c\geq0$ implies that $c$ can be performed. 
A consequence claim $s$ can be performed if the related primary claim $c$ is not performed,
expressed by $\Phi_s\equiv\phi_{\owner}\wedge (d_c=-1)\wedge\phi_c\wedge\phi_s\wedge d_s\geq0$.
Analysis II determines whether an execution of the considered SPA exists by 
assessing the satisfiability of $\phi_\SPA^B \wedge \phi_{\textit{soft}}^B$.

The SMT solver invoked by \textit{ContractCheck} will determine whether 
the constraint system $\phi_\SPA \wedge \phi_{\textit{soft}}^B$ is satisfiable,
in which case a satisfying variable assignment will be returned.
If it is unsatisfiable, the SMT solver returns an unsatisfiability core, 
which is a minimal subset of the analyzed assertions 
that contradicts satisfiability.
This subset indicates which claims in the SPA contradict another
and, hence, causes an inconsistency in the SPA. 

\paragraph{Analysis I} 
The SMT solver analyzes the formula 
$\Phi_{\textit{TransferClaim}} \equiv
\phi_\owner^B \; \wedge \; \phi_{\textit{TransferClaim}} \wedge\;
d_{u}\geq 0,$
which can equivalently be rewritten as
$owner(\textit{Bakery}) = \textit{Bank} \wedge
(d_u=-1 \vee (28\leq d_u 
\Rightarrow \owner(\textit{Bakery})=\textit{Eva})) \; \wedge \; d_u\geq 0.$
Using functional consistency reasoning, the SMT solver recognizes the 
contradiction between $\owner(\textit{Bakery})=\textit{Bank}$ and
$\owner(\textit{Bakery})=\textit{Eva}$. \textit{ContractCheck}
reports this inconsistency to the user by identifying the blocks that
are in contradiction.

For $\Phi_{\textit{PayClaim}}$, \textit{ContractCheck} finds a satisfying assignments
with $d_z=28$ and $\owner(\textit{Price})=\textit{Chris}$, which means
all premises for performing this claim are satisfied. Likewise, for 
$\Phi_{\textit{PretzelWarranty}}$ a satisfying model with $d_g = -1$ and
$\mathit{pretzels}=10000$ is computed, which indicates that the warranty
claim does not need to be performed.

\textit{ContractCheck} determines $\phi_{\textit{Claim1}}$ to be unsatisfiable,
which entails that the time constraints are consistent. 
We now consider a simplified version of the compensation claim Claim2:
$\phi'_{Claim2} \equiv \phi_{SPA} \wedge \mathtt{Claim2.Limitation} < 
\mathtt{Claim2.DueDate}$. In it, \texttt{Claim2.Limitation} denotes the time 
limit after signing from which on no further claims due to the SPA 
exist and has the value $70$. 
The \texttt{Claim2.Limitation} is computed as $d_w + 28 + 14$, 
where $d_w$ is the limit after which the warranty claim needs to be
performed. \textit{ContractCheck} computes an unsatisfiability core
with $d_w = 29$, which means that Eva can wait to perform the 
warranty until no further obligations from the contract can be 
enforced. The result of this analysis is displayed to the user 
as depicted in Figure~\ref{fig:claim2}.

{\centering
\begin{figure}
\begin{subfigure}{.5\textwidth}
  \centering
  \includegraphics[width=.7\linewidth]{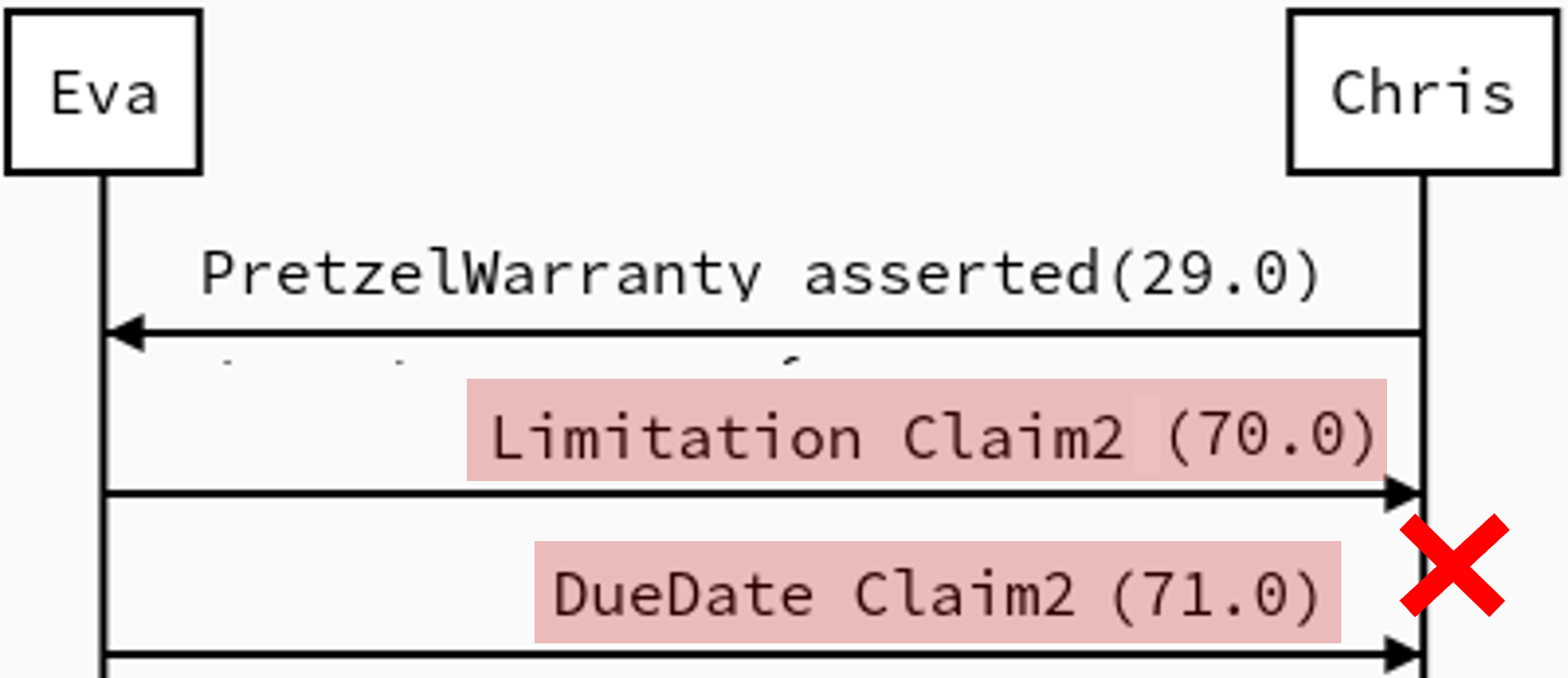}
  \caption{Non-performability of the TransferClaim in the Bakery SPA.}
  \label{fig:claim2}
\end{subfigure}%
\begin{subfigure}{.5\textwidth}
  \centering
  \includegraphics[width=.7\linewidth]{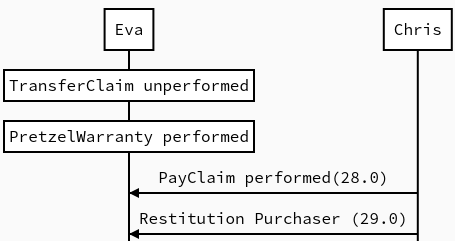}
  \caption{SPA contract execution.}
  \label{fig:result_seq}
\end{subfigure}%
\end{figure}
}

\paragraph{Analysis II} We analyze whether an execution of 
the SPA exists by assessing the satisfiability of 
$\Phi_\SPA\equiv\phi_\SPA\wedge \phi_{\textit{soft}}.$
\textit{ContractCheck} returns a satisfiable model with 
the following variable assignments: $d_u = -1$ (the bakery is not 
transferred), $d_z = 28$ (payment of the purchase prize has been made),
$d_g = -1$ (no warranty claim performed), $d_n = -1$ (not pretzel
consequence claim performed), $d_s=-1$ (no compensation payment), 
$d_{\mathit{RestitutionSeller}} = -1$ (no restitution for the seller) and
$d_{\mathit{RestitutionPurchaser}} = -1$ (restitution of the purchaser on
day 29). This shows that an inconsistent SPA, in particular one that has a  
non-performable primary claim, can nonetheless be executed. 
The analysis result is displayed to the user as depicted in Figure~\ref{fig:result_seq}.

\section{Research Directions}

There are various ways in which we plan to extend the current set of analyses
implemented in \textit{ContractCheck}.
First, we plan to define a dynamic, concurrent and state-based execution model for 
the execution of a contract, in particular an SPA. 
Notice that we currently only encode the premises for the performance of a claim,
but not the effect on its state, and consequently no state changes.
The state machine representing the state changes of a  
contract claim will be based on the one sketched in Figure~\ref{fig:claim}. 
Second, we will incorporate an idea of the contractual framework vs.\ the environment.
In the SPA we have encoded the fact that ownership of the bakery is transferred to the 
bank as a clause in the contract, whereas this is a fact that lies in the environment.
Third, we will add state-based model components that represent the actors involved, 
i.e., the seller and the purchaser. These state models, which can be probabilistic, 
represent the decisions that the actors take in terms of the performance of claims.
Fourth, when adding notions of rewards or costs to these models, we can then 
develop algorithmic techniques that 
address the question what contract situations and which 
contract execution strategies can be beneficial or advantageous for the one or the
other contract party. 
These analyses will rely on reward models and game
theoretical approaches and will pave the way for the development of an advisory tool.
Fifth, we will extend the set of blocks that we have defined so far to accommodate
more complex contracts. Sixth, we will explore ways to directly link the textual form
of an SPA contract to its logical representation by exploiting techniques from 
from Automated Document Analysis and Natural Language Processing.

\section{Conclusions}
We have presented the analysis of SPAs in the context of the  \textit{ContractCheck} tool
that maps logical constraints expressed in 
natural language contract text via a structure 
language representation to decidable fragments of first-order logic.
The constraints are then checked for consistency using SMT solving. 
The degree of automation achieved by using SMT solving will greatly 
facilitate the incorporation of this methodology into practical 
contract generation tools.
We have finally illustrated the analysis using a simple example, and have
proposed directions for future research.

\pagebreak

\bibliographystyle{ACM-Reference-Format}
\bibliography{bib}

\end{document}